\documentclass[aps,pra,twocolumn,superscriptaddress]{revtex4}
\usepackage{amsfonts}
\usepackage{amssymb}
\usepackage{amsmath}
\usepackage{epsfig}
\usepackage{color}
\usepackage{graphics, graphicx}
\usepackage{bbold}
\usepackage{psfrag}
\usepackage{mathcomp}
\usepackage{subfigure}
\usepackage{verbatim}
\usepackage{color}
\usepackage[colorlinks,citecolor=blue]{hyperref}

\usepackage{soul}
\usepackage{ulem}
\usepackage{cancel}

\begin{document}

\title{Interaction-induced phase transitions at topological quantum criticality of an extended Su-Schrieffer-Heeger model}

\author{Xiaofan Zhou}
\affiliation{State Key Laboratory of Quantum Optics and Quantum Optics Devices, Institute of Laser spectroscopy, Shanxi University , Taiyuan 030006, China}
\affiliation{Collaborative Innovation Center of Extreme Optics, Shanxi University, Taiyuan, Shanxi 030006, China}
\author{Suotang Jia}
\affiliation{State Key Laboratory of Quantum Optics and Quantum Optics Devices, Institute of Laser spectroscopy, Shanxi University, Taiyuan 030006, China}
\affiliation{Collaborative Innovation Center of Extreme Optics, Shanxi University, Taiyuan, Shanxi 030006, China}
\author{Jian-Song Pan}
\email{panjsong@scu.edu.cn}
\affiliation{College of Physics, Sichuan University, Chengdu 610065, China}
\affiliation{Key Laboratory of High Energy Density Physics and Technology of Ministry of Education, Sichuan University, Chengdu 610065, China}

\begin{abstract}
Topological phases at quantum criticality attract much attention recently. Here we numerically study the interaction-induced phase transitions at around the topological quantum critical points of an extended Su-Schrieffer-Heeger (SSH) chain with next-nearest-neighbor hopping. This extended SSH model shows topological phase transitions between the topologically trivial and nontrivial critical phases when interaction is absent. So long as the interaction terms are turned on, the topologically nontrivial (trivial) critical phases are driven into topologically nontrivial (trivial) insulator phases with finite energy gaps. Particularly, we find the trivial insulator phase is further driven to the nontrivial insulator phase, through interaction-induced topological phase transition, although interaction generally is harmful to nontrivial topology. The stability of trivial insulator phase against interaction tends to vanish at the multicritical point that separates the trivial and nontrivial critical phases. Our work provides a concrete example for manifesting the impact of interaction on topological quantum criticality.
\end{abstract}

\maketitle

\section{Introduction}
\label{Introduction}
The concept of topological phase has taken condensed matter physics by storm in the past decades~\cite{Kane2005,Bernevig2006}. The key feature of topological phase lies on that its interior is insulating while its surface is conducting~\cite{Hasan2011}. The surface conductance is quantized in general, which is essentially protected by topological invariant defined on the energy bands below the Fermi surface~\cite{qi2011topological}. In this scenario, an energy gap upon the Fermi surface is necessary for defining the topological invariant. Even for interacting systems, energy gap upon the ground state generally is also required for defining the symmetry-protected topological phases~\cite{gu2009tensor,pollman2012symmetry}. A widely believed wisdom is that the topological edge modes are protected by the energy gap upon the ground state and thus are unstable when the bulk energy gap is closing.

More and more studies show that edge modes even can survive at gapless quantum criticality~\cite{kestner2011prediction,cheng2011majorana,fidkowski2011majorana,sau2011number}. Nevertheless, the exponential localization are still associated with gapped degrees of freedom (i.e., there are exponentially decaying correlation functions). R. Verresen \emph{et al.} demonstrate the gapless edge modes at quantum criticality of topological Fermi chain (e.g., topological chain with time-reversal symmetry) are essentially different since they even have no gapped degrees of freedom~\cite{verresen2018prediction}. These gapless phases with localized edge modes are also protected by topological invariants. Phase transitions between different critical topological phases occur at multicritical points, as exemplified with the critical chains with next nearest coupling terms ~\cite{kumar2023signatures}. Although the role of interactions can be abstractly included in the framework of symmetry-enriched quantum criticality, similar to the interacting symmetry-protected topological phases~\cite{verresen2021gapless,verresen2020topology}, studies with specific example are still rare.

In this paper, we systematically studied the impacts of interactions in an extended SSH chain with next-nearest-neighbor hopping terms based on the density-matrix renormalization-group (DMRG) method~\cite{dmrg1,dmrg2}. We map the phase diagram by calculating different physical quantities, including entanglement spectra, entanglement entropy, winding number and energy gap. The free limit of this model shows phase transition between the topologically nontrivial and trivial critical phases. The phase diagram shows that, interaction opens energy gap and drives the topologically nontrivial (trivial) critical phases into the topological (normal) insulator phases immediately. In general, interaction is an antagonist in the phase diagram of topological model, i.e., it usually tends to expand the trivial phases. Indeed, we find the nontrivial insulator phase can be further broken by interaction, i.e., became trivial insulator phase through an interaction-induced topological phase transition. However, we find the trivial insulator phase can also be driven into the nontrivial insulator phase by interaction. The trivial insulator phase is very unstable at around the multicritical point separating the topologically nontrivial and trivial critical phases. The stability of trivial insulator phase even vanishes toward the multicritical point. The phase boundary between the trivial and nontrivial insulator phases emanated from the multicritical point thus bends approach the parameter line of trivial critical phase.

The rest of this paper is organized as follows. We introduce the considered model in Sec. II. The definitions of quantities employed to characterize the phase transitions are introduced in Sec. III. Many-body phase diagram is presented in Sec. IV. The interaction-induced topological phase transitions are characterized in detail in Sec. V and VI. A brief summary is given in Sec. VII.

\section{Model and Hamiltonian}
\label{Model and Hamiltonian}

\begin{figure}[t]
\centering
\includegraphics[width = 3.5in]{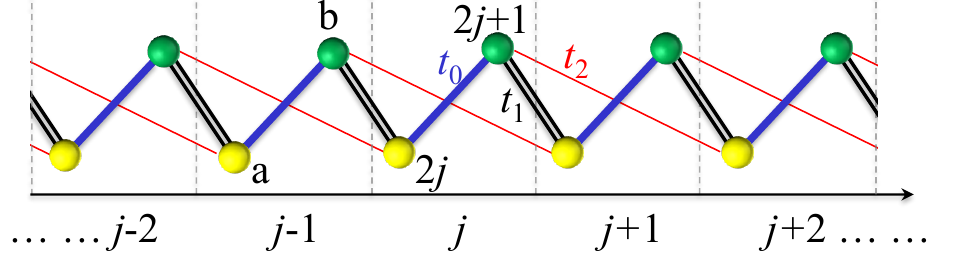} \hskip 0.0cm
\caption{The sketch of the extended SSH model with next-nearest-neighbor hopping terms. Here $j$ denotes the index of unit cell, $i=2j, 2j+1$ represents the index of lattice site, and a and b label the two sublattices.}
\label{fig:sketch}
\end{figure}

The model we employ to show the impact of interaction on critical phases is the extended SSH model with next-nearest-neighbor hopping terms, as illustrated in Fig.~\ref{fig:sketch}. The Hamiltonian is given by
\begin{eqnarray}
H \!\! &=& \!\! \sum_j (t_0 c_{j,a}^{\dag}\hat{c}_{j,b} +  t_1 \hat{c}_{j+1,a}^{\dag}\hat{c}_{j,b}
+ t_2 \hat{c}_{j+2,a}^{\dag}\hat{c}_{j,b} +\text{H.c.}) \notag
\\
&&+ V\sum_j (\hat{n}_{j,a} \hat{n}_{j,b}+\hat{n}_{j,b} \hat{n}_{j+1,a}),
\label{H1}
\end{eqnarray}
where $\hat{c}_{j,a}=\hat{c}_{2j}$ ($\hat{c}_{j,b}=\hat{c}_{2j+1}$) is the fermionic annihilation
operator at the $j$-th cell of sublattice a (b), $\hat{n}_{j,a(b)}=\hat{c}_{j,a(b)}^{\dag }%
\hat{c}_{j,a(b)}$ are the density operators,
$t_{n=0,1}$ denote the stagger nearest-neighbor hopping coefficients, $t_{2}$ measures the next-nearest-neighbor hopping strength, and $V$ characterizes the density-density interaction strength. Here we assume the difference between the intra cell and the inter cell interaction strengths is ignorable for simplicity.

The non-interacting phase diagram of the long-range SSH model not only displays the topologically trivial and nontrivial critical phases, but also shows clear phase transition between these critical phases~\cite{kumar2023signatures}. Specifically, the free limit of model (\ref{H1}) with $V=0$ is in the non-trivial (trivial) critical phase when $|t_2|>|t_1|$ ($|t_2|<|t_1|$) at the quantum criticality $t_{1}=\pm(t_{0}+t_{2})$, which separates different gapped insulator phases. For fixed $t_{0}$, the trivial and nontrivial phases are separated by the multicriticalities $t_2 =\pm t_0$. The non-trivial critical phase hosts stable edge modes, although the bulk gap is vanishing.

In the following, we focus on the critical phases on $t_{2}$ axis by keeping $t_1 = t_2+t_0$ and $t_0=1$ (the energy unit). The phase diagram is an interacting extension of the critical phases to $t_{2}-V$ plane. The number of cells is set up to 240, corresponding to a total site of $L=480$. Half filling with atom number $N=L/2$ is adopted. We retain 400 truncated states per DMRG block and perform 30 sweeps with acceptable
truncation errors.

\section{Physical quantities employed for characterizing the phase diagram}
In this section, we introduce the definitions of several quantities employed to characterize the topological phases and topological phase transitions, including entanglement spectrum, entanglement entropy, winding number, density distribution of edge modes and energy gap.
The entanglement spectrum is given by the logarithmic function of the Schmidt
values~\cite{Li2008},
\begin{equation}
\xi _{m}=-\ln (\rho _{m}),
\end{equation}%
where $\rho_{m}$ is the $m$-th eigenvalue of the reduced density matrix $\hat{%
\rho}_{l}=\mathrm{Tr}_{L-l}|\psi \rangle \langle \psi |$, with the ground-state wave-function of Hamiltonian (\ref{H1}) $|\psi \rangle
$. Here $l$ is the length of the left block for a specific bipartition.  The system is topologically nontrivial provided the entanglement spectrum is degenerate, since the entanglement spectrum is associated with the energy spectrum of edge excitations \cite{Li2008,Zhao2015,Yoshida2014,Turner2011,Pollmann2010,Fidkowski2010,Flammia2009,Yu2024}.

The quantum criticality of
interaction-driven topological phase transition can be characterized with the von
Neumann entropy \cite%
{Flammia2009,Hastings2010,Daley2012,Abanin2012,jiang2012,Islam2015}
\begin{equation}
S_{\mathrm{vN}}=-\mathrm{Tr}_{l}[\hat{\rho}_{l}\log \hat{\rho}_{l}],
\end{equation}%
with $l=L/2$. It is believed that the property underlying the long-range correlations is
entanglement~\cite{osborne2002entanglement}, and, on the other hand, the correlation length becomes divergent at the critical point of continuous phase transition~\cite{SachdevQPT}. The divergence of von Neumann entropy at the critical point thus indicates a continuous transition~\cite{Pollmann2010}. Besides, the von Neumann entropy also reveals the central charge of the conformal field
theory underlying the critical behavior, which
generally determines the effective field theory and reflects the universality class of phase transition~\cite{pasquale}.
For a critical system under open boundary condition, the von Neumann entropy of a sub-chain of length $l$ scales as
\begin{equation}
S_{\mathrm{vN}}(l)=\frac{c}{6}\ln \left[ \sin \frac{\pi l}{L}\right] +\text{const},
\end{equation}%
in which, the slope at large distance gives the central charge $c$ of the
conformal field theory~\cite%
{centralcharge1,centralcharge2,centralcharge3,centralcharge4}.

The density distribution of the edge modes can be calculated as
\begin{equation}
\langle \Delta \hat{n}_i \rangle = \langle \hat{n}_i(N+1) \rangle - \langle%
\hat{n}_i(N)\rangle,
\end{equation}%
with $\hat{n}_{i}=\langle\hat{c}_{i}^{\dagger}\hat{c}_{i}\rangle$, and $\langle\hat{n}_i(N)\rangle$ the density distribution for $N$ atoms
under the open boundary condition.

We also calculate the spin correlation function $S(k)=S_{x}(k)\sigma_{x}+S_{y}(k)\sigma_{y}+S_{z}(k)\sigma_{z}$, where $S_{x}(k)=\langle\hat{c}_{ka}\hat{c}_{kb}^{\dagger}+\hat{c}_{kb}\hat{c}_{ka}^{\dagger}\rangle/2$, $S_{y}(k)=i\langle\hat{c}_{ka}\hat{c}_{kb}^{\dagger}-\hat{c}_{kb}\hat{c}_{ka}^{\dagger}\rangle/2$ and $S_{z}(k)=\langle\hat{c}_{ka}\hat{c}_{ka}^{\dagger}-\hat{c}_{kb}\hat{c}_{kb}^{\dagger}\rangle/2$, with $\hat{c}_{k,a,b}=L^{-1/2}\sum_{j}\hat{c}_{j,a,b}e^{-ikj}$.  In the regime where the chiral symmetry has not been spontaneously broken by interaction terms (it is spontaneously broken in CDW phase),  $S_{z}(k)=0$ and a winding number for the spin texture can be extracted from
\begin{equation}
\omega = \frac{\phi_{\pi-\delta}-\phi_{-\pi+\delta}}{2\pi}
\end{equation}%
where $\phi_{k}=\arctan (S^y_k / S^x_k) $, and the infinitesimal quantity $\delta$ is employed to exclude the singular points $k=\pm \pi$ in the gapless critical phases. The multivalued function $\phi_{k}$ is restricted to be continuous except at the singular points. In principle, the TI phase (trivial BI phase) has a winding number $w=1$ ($w=0$). In contrast, the winding number of the nontrivial (trivial) free critical phase is $w=3/2$ and $1/2$, respectively~\cite{verresen2018prediction}.

%As we all known, the nearest-neighbor interaction $V$ can induce the CDW phase, in which the CDW order parameters can be defined as
%\begin{equation}\label{CDW_order}
%C= \frac{1}{L} \sum_{i=1}^{L} (-1)^i \langle \hat{n}_i \rangle.
%\end{equation}

%The energy gap can be defined as,
%\begin{equation}
%\Delta = \mu^{+} - \mu^{-},
%\end{equation}%
%in which, $\mu^{+} = E_g(N+1) - E_g(N)$, $\mu^{-} = E_g(N) - E_g(N-1)$.
%$E_g(N)$ is the ground state energy for $N$ atoms.

Besides, we also calculate the energy gap $\Delta$, which typically vanishes at the topological transition points. In the topologically nontrivial phases,  $\Delta=\mu_{L/2+2}-\mu_{L/2-1}$ in which, the chemical potential $\mu_{N} = E_N - E_{N-1}$ with $N$-particle ground-state energy $E_{N}$, since the pair of degenerate zero-energy modes are occupied for the $L/2$- and $(L/2+1)$-particle ground states. In the trivial phases, $\Delta=\mu_{L/2+1}-\mu_{L/2}\approx \mu_{L/2+2}-\mu_{L/2-1}$, as the zero-energy modes merge into the bulk.

\begin{figure}[t]
\centering
\includegraphics[width = 3.5in]{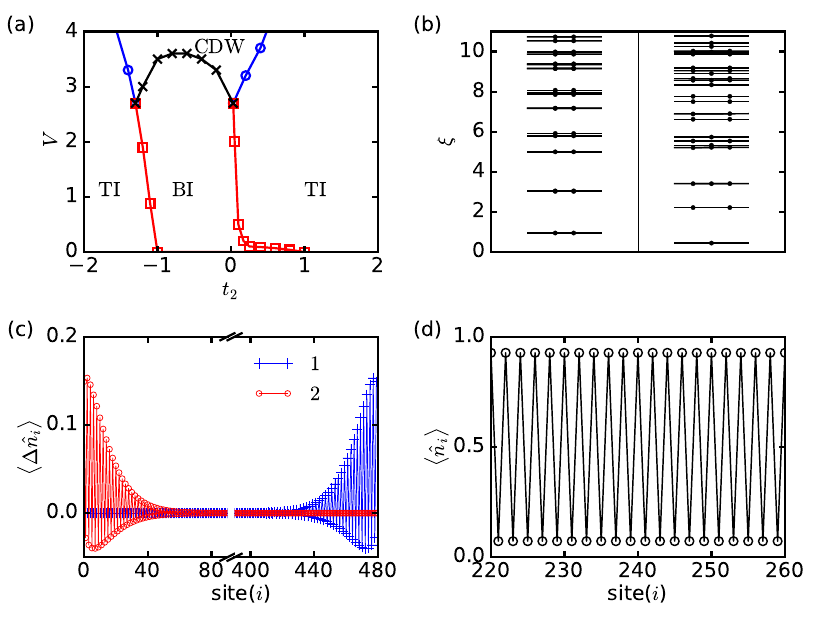} \hskip 0.0cm
\caption{(a) The phase diagram on the $t_2-V$ plane of thermodynamic limit $L\rightarrow \infty$.
(b) The entanglement spectrum $\protect\xi$ with $t_2=-1.1$, $V=0.2$ (left TI) and $V=0.8$ (right BI), and $L=480$.
(c) The density distribution of the two-fold degenerate edge modes of TI phase with $t_2=1.0$, $V=2.5$, and $L=480$.
(d) The atomic density distribution near the center lattice of CDW phase with $t_2=-0.5$, $V=4.0$, and $L=480$.}
\label{fig:phase diagram spinless}
\end{figure}

\section{phase diagram}
\label{phase diagram}

Based on the calculated degeneracy of entanglement spectrum, density distribution as well as the finite-size scaling, we map the phase
diagram on the $t_2-V$ plane of thermodynamic limit, as shown in Fig.~\ref{fig:phase diagram spinless}(a). The topological phases are also consistent with the calculation of winding number. The phase boundaries are also confirmed by the calculations of entanglement entropy and energy gap. Basically, besides the critical phases, the phase diagram contains three new phases: topological insulator (TI), band insulator (BI) with trivial topological invariant and charge density wave (CDW) phases.

The nontrivial/trivial phases display degenerate/nondegenerate entanglement spectrum, as shown in Fig.~\ref{fig:phase diagram spinless}(b). All levels of $\protect\xi$ of nontrivial phase for weak interaction are two-fold degenerate. However, some $\protect\xi$ is no longer degenerate for strong interaction strength.
The TI phase has topologically protected edge states under open boundary
conditions, which are two-fold degenerate, and only one edge mode is occupied on one edge side for each degenerate ground state, as shown in Fig.~\ref{fig:phase diagram spinless}(c).
For large nearest-neighbor interaction strength $V$,
the density distribution of the ground state always modulates
along real lattice space with a period of 2, with the corresponding phase CDW, which is shown in Fig.~\ref{fig:phase diagram spinless}(d).

From the phase diagram in Fig.~\ref{fig:phase diagram spinless}(a), we can find interaction drive the topologically trivial (nontrivial) critical phases on $|t_{2}|<1$ ($|t_{2}|>1$) to the topologically trivial BI phase (nontrivial TI phase), respectively. The vanishing energy gap at the critical phases becomes finite in the BI and TI phases. This observation is different from the example presented in Ref.~\cite{verresen2018prediction}, where the topological critical phase is unchanged in the presence of the interaction of a special form.

When interaction becomes strong enough, both TI and BI phases are all driven into the CDW phase. This observation is very natural, because our interaction include repulsive nearest-neighbor interaction. The density distribution in the CDW phase shows staggered structure, as shown in Fig.~\ref{fig:phase diagram spinless}(d). The chiral symmetry, which protects the topological invariant in TI phase, is spontaneously broken. The topological invariant defined for characterizing the TI and BI phases is thus no longer applicable for the CDW phase~\cite{zhou2023exploring}.

The most intriguing discovery is that, the trivial BI phase at around the multicritical point $t_{2}=1$ is unstable, and is driven into TI phase immediately. The BI phase thus looks like a wedge plugging to the multicritical point $t_{2}=1$. In principle, the energy gap is opened and immediately closed, is opened again when increasing the interaction strength. This discovery not only implies the BI phase is very unstable at around the multicritical points $t_{2}=1$ in terms of interaction, but also implies interaction may play multiple roles for the topological phases. In the following section, we present more details of these interacting-induced topological phases and phase transitions by depicting the variations of different quantities along the vertical cutting lines by increasing the interaction strength on the phase diagram.

\begin{figure}[t]
\centering
\includegraphics[width = 3.5in]{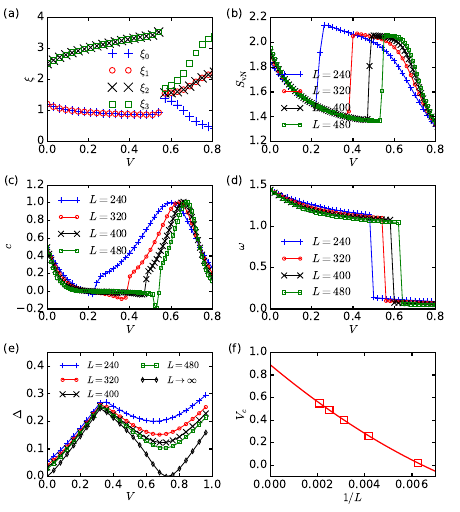} \hskip 0.0cm
\caption{Typical interaction-induced phase transition at nontrivial quantum criticality.
(a) the lowest four levels in the entanglement spectrum $\protect%
\xi_{m}$ ($m=0,1,2,3$) of $L=480$,
(b) The von Neumann entropy $S_{\mathrm{vN}}$,
(c) the central charge $c$,
(d) the winding number $\omega$,
(e) the energy gap $\Delta$,
as functions of the interaction strength $V$ with $t_2=-1.1$ for several sizes.
(f) The finite-size scaling of the critical points.}
\label{fig:phase_transition_11V}
\end{figure}

\section{Interaction-induced topological phase transition at nontrivial quantum criticality}
To elaborate the interaction-induced topological phase transitions at nontrivial quantum criticality, we depict the variation of  entanglement spectrum $\xi$ (a), entanglement entropy $S_{\mathrm{vN}}$ (b), central charge $c$ (c), winding number $w$ (d) and energy gap $\Delta$ (e) along the vertical line at $t_2=-1.1$, and the finite-size scaling of the critical interaction (f) in Fig.~\ref{fig:phase_transition_11V}. Due to the lack of chiral symmetry and thus the absence of topological invariant in the CDW phase, we only focus on the part of parameter without CDW phase. On the whole, point $V=0$ does not show sharp changes of those quantities. Instead, the interaction-induced topological phase transitions between the TI and BI phases at finite interaction strengths manifests them with clear critical behaviors.

From Fig.~\ref{fig:phase_transition_11V}(a), we can find the entanglement spectrum $\xi$ is degenerate (non-degenerate) in TI (BI) phase, which is a smoking gun for the interaction-induced phase transition between TI and BI phases. The finite-size scaling of energy gap closes and re-opens at the topological phase transition point, as shown in Fig.~\ref{fig:phase_transition_11V}(e). The finite-size scaling of the critical interaction for the TI-BI phase transition is shown in Fig.~\ref{fig:phase_transition_11V}(f). The thermal thermodynamic limit of the critical interaction locates at $V_{c}\approx0.88$.

In Fig.~\ref{fig:phase_transition_11V}(b) with $t_{2}=-1.1$, where the non-interacting limit is in the trivial critical phases, the entanglement entropy $S_{\mathrm{vN}}$ continuously decreases when turning on the interaction terms. There exists a critical interaction where $S_{\mathrm{vN}}$ suddenly jumps, which is just the phase transition between TI and BI phases. The central charge $c$ and winding number $w$ also show similar features from Figs.~\ref{fig:phase_transition_11V}(c) and \ref{fig:phase_transition_11V}(d).

The central charge starts from $0.5$ and the winding number starts from $1.5$, which is consist with the analysis without interaction~\cite{kumar2023signatures}. They also continuously decreases as increasing interaction. In principle, the winding number should be $1$ in the TI phase. That the continuous deviation of winding number from $1.5$, instead of the sudden jump to $1$, may be due to the slow increase of the energy gap. The winding number finally drops to a small value approaching zero in the BI phase.

\begin{figure}[t]
\centering
\includegraphics[width = 3.5in]{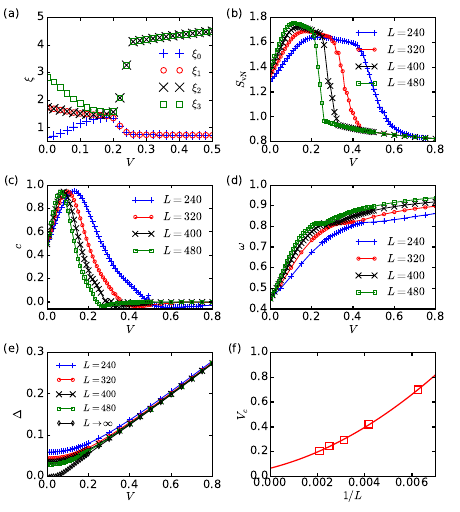} \hskip 0.0cm
\caption{Typical interaction-induced phase transition at trivial quantum criticality.
(a) the lowest four levels in the entanglement spectrum $\protect%
\xi_{m}$ ($m=0,1,2,3$) of $L=480$,
(b) The von Neumann entropy $S_{\mathrm{vN}}$,
(c) the central charge $c$,
(d) the winding number $\omega$, (e) the energy gap $\Delta$,
as functions of the interaction strength $V$ with $t_2=0.25$ for several sizes.
(f) The finite-size scaling of the critical points.}
\label{fig:phase_transition_25V}
\end{figure}

\section{Interaction-induced topological phase transition at trivial quantum criticality}
Parallel discussions on an interaction-induced topological phase transition at trivial quantum criticality are presented in this section. The variations of different quantities in the $V$ direction at a typical trivial quantum criticality $t_{2}=0.25$ are shown in Fig.~\ref{fig:phase_transition_25V}. From Fig.~\ref{fig:phase_transition_25V}(a), we can find the entanglement spectrum $\xi$ is non-degenerate (degenerate) in BI (TI) phase, which is a smoking gun for the interaction-induced phase transition between TI and BI phases. The finite-size scaling of the critical interaction for the TI-BI phase transition is shown in Fig.~\ref{fig:phase_transition_25V}(f).

The entanglement entropy $S_{\mathrm{vN}}$ in Fig.~\ref{fig:phase_transition_25V}(b) continuously decreases when turning on the interaction terms. $S_{\mathrm{vN}}$ also suddenly jumps at around the topological transition point between BI and TI phases. The central charge $c$ and winding number $w$ also show similar features from Figs.~\ref{fig:phase_transition_25V}(c) and \ref{fig:phase_transition_25V}(d). The central charge starts from $0.5$ and the winding number starts from $0.5$, which is also consist with the analysis without interaction~\cite{kumar2023signatures}. They also continuously decreases as increasing interaction. In principle, the winding number should be $1$ in the TI phase. That the continuous deviation of winding number from $0.5$, instead of the sudden jump to $0$, may be due to the small energy gap. The winding number finally increases to a value approaching $1$ in the TI phase. The thermal thermodynamic limit of the critical interaction locates at $V_{c}\approx0.067$. The typical close-reopening behavior of energy gap at topological phase transition is unclear in this case, as shown in Fig.~\ref{fig:phase_transition_25V}(e). It may be due to the critical interaction strength is too small, and thus the energy gap in the BI phase is too small. As a result, the energy gap almost always at around zero in the BI phase with $V\in (0, V_{c})$.

%The phase transition as function $V$ with $t_2=-1.6$, as shown in Fig.~\ref{fig:phase_transition_V}.
%\begin{figure*}[t]
%\centering,
%\includegraphics[width = 7.0in]{phase_transition} \hskip 0.0cm
%\caption{(a) The lowest four levels in the entanglement spectrum $\protect%
%\xi_{i}$ ($i=0,1,2,3$),
%(b) the von Neumann entropy $S_{\mathrm{vN}}$,
%(c) the CDW order parameter $C$,
%(d) the central charge $c$,
%(e) energy spectrum $\mu$,
%(f) fidelity $F(\psi^{'},\psi)$
%as functions of the interaction strength $V$ with $t_2=-1.6$. }
%\label{fig:phase_transition_V}
%\end{figure*}

%The phase transition as function $t_2$, with the interaction strength $V=2.0$, as shown in Fig.~\ref{fig:phase_transition_t2}.

%\begin{figure}[t]
%\centering
%\includegraphics[width = 1.2in]{ES_t2_V=2.000000_L=160_N=80_o.eps} \hskip 0.0cm
%\includegraphics[width = 1.2in]{Entropy_t2_V=2.000000_L=160_N=80_o.eps} \hskip 0.0cm
%\includegraphics[width = 1.2in]{c_t2_V=2.000000_L=160_N=80_o.eps} \vskip 0.0cm
%%
%\includegraphics[width = 1.2in]{ES_t2_V=2.000000_L=160_N=80_p.eps} \hskip 0.0cm
%\includegraphics[width = 1.2in]{Entropy_t2_V=2.000000_L=160_N=80_p.eps} \hskip 0.0cm
%\includegraphics[width = 1.2in]{c_t2_V=2.000000_L=160_N=80_p.eps} \hskip 0.0cm
%\caption{(a) The lowest four levels in the entanglement spectrum $\protect%
%\xi_{i}$ ($i=0,1,2,3$),
%(b) the von Neumann entropy $S_{\mathrm{vN}}$,
%(c) the  energy gap $\Delta$,
%(d) the CDW order parameter $C$,
%as functions of $t_2$ with the interaction strength $V=2.5$.}
%\label{fig:phase_transition_t2}
%\end{figure}

\section{Conclusions}
\label{Conclusions}

In conclusion, we numerically study the interaction-induced topological phase transitions at quantum criticality of the long-range SSH models with DMRG method. On the topologically nontrivial (trivial) quantum criticality, the interaction terms open energy gap and induce topologically nontrivial (trivial) insulator phases first. However, by increasing the interaction strength, the nontrivial (trivial) insulator phases can be further driven to cross a interaction-induced topological phase transition to the trivial (nontrivial) insulator phases. We would like to emphasize that, the interaction-induced trivial-to-nontrivial insulator phase transition has been rarely reported. Our work unveils the interplay between interaction and quantum criticality and thus may inspire more research interest on interacting topological critical phases.

\section*{Acknowledgments}
We thank Prof. Jin Zhang from Chongqing University for the helpful discussions. X.Z. and S.J. are supported by
National Key R\&D Program of China under Grant No.
2022YFA1404003, the National Natural Science Foundation of China (NSFC) under Grant No.~12004230, 12174233 and 12034012, the Research Project Supported by Shanxi Scholarship Council of China and Shanxi '1331KSC'. J.S.P. is supported by the Science Specialty Program of Sichuan University under Grant No. 2020SCUNL210 and the Fundamental Research Funds for the Central Universities under Grant No. 20826041G4165.
The DMRG calculation is taken with the ALPSCore library~\cite{ALPS}, which is constructed upon the original ALPS project~\cite{ALPS2}.

\end{document}